\documentclass[aps,prl,reprint,superscriptaddress,footinbib,longbibliography]{revtex4-1}

\usepackage[version=4]{mhchem}
\ExplSyntaxOn
\keys_define:nn { mhchem }
 {
  arrow-min-length .code:n =
   \cs_set:Npn \__mhchem_arrow_options_minLength:n { {#1} } % default is 2em
 }
\ExplSyntaxOff
\mhchemoptions{arrow-min-length=1em}
\usepackage{tikz}
\usepackage{csquotes}
\usepackage{xcolor}
\usepackage{booktabs}
\usepackage{pifont}
\usepackage{xr}	
\usepackage[capitalize]{cleveref}	% ! Must come after xr package
\crefname{section}{section}{sections}

\makeatletter
\newcommand*{\addFileDependency}[1]{% argument=file name and extension
  \typeout{(#1)}
  \@addtofilelist{#1}
  \IfFileExists{#1}{}{\typeout{No file #1.}}
}
\makeatother

\newcommand*{\myexternaldocument}[1]{%
    \externaldocument{#1}%
    \addFileDependency{#1.tex}%
    \addFileDependency{#1.aux}%
}
\myexternaldocument{sm}

\begin{document}
% %TC commands are macros for TeXcount. We use them to remove whatever part which should not be counted.
% NB: OverLeaf (which uses TeXcount) does not show captions word count, so it is better to download the project and run texcount locally.

%TC:ignore
\title{Charge transfer of polyatomic molecules in ion-atom hybrid traps: Stereodynamics in the millikelvin regime}
\author{Alexandre Voute}
\email[Corresponding author: ]{alexandre.voute@universite-paris-saclay.fr}
\affiliation{Universit\'e Paris-Saclay, CNRS, Institut des Sciences Mol\'eculaires d'Orsay, 91405, Orsay, France.}
\affiliation{Universit\'e Paris-Saclay, CNRS, Laboratoire Aim\'e-Cotton, 91405, Orsay, France.}
\author{Alexander D\"{o}rfler}
\affiliation{Department of Chemistry, University of Basel, Klingelbergstrasse 80, 4056 Basel, Switzerland}
\author{Laurent Wiesenfeld}
\author{Olivier Dulieu}
\affiliation{Universit\'e Paris-Saclay, CNRS, Laboratoire Aim\'e-Cotton, 91405, Orsay, France.}
\author{Fabien Gatti}
\author{Daniel Pel\'{a}ez}
\affiliation{Universit\'e Paris-Saclay, CNRS, Institut des Sciences Mol\'eculaires d'Orsay, 91405, Orsay, France.}
\author{Stefan Willitsch}
\email[Corresponding author: ]{stefan.willitsch@unibas.ch}
\affiliation{Department of Chemistry, University of Basel, Klingelbergstrasse 80, 4056 Basel, Switzerland}

\date{\today}

\begin{abstract}
Rate constants for the charge transfer reaction between \ce{N2H+} and \ce{Rb} in the mK regime are measured in an ion-atom hybrid trap and are found to be lower than the Langevin capture limit. Multireference ab initio computation of the potential energy surfaces involved in the reaction reveals that the low-temperature charge transfer is hindered by short-range features highly dependent on the collision angle and is promoted by a deformation of the molecular frame. The present study highlights the importance of polyatomic effects and of stereodynamics in cold molecular ion-neutral collisions.
\end{abstract}


\maketitle

%%%%%%%%%%%%%%%%%%%%%%%%%%%%%%%%%%%%%%%%%%%%%%%%%%%%
%TC:endignore
\emph{Introduction.}
Over the past decade, the advent of cold ion-atom hybrid trapping experiments has opened up a range of new research avenues in both physics and chemistry \cite{tomza19a}.
These include the exploration of new cooling mechanisms \cite{zipkes10a,ravi12a,mahdian21a,rellergert13a},
the study of unusual light-driven reactive processes \cite{hall11a, mills19a},
the characterization of many-body dynamics \cite{kruekow16a}, and the investigation of ion-atom interactions in the quantum regime \cite{meir16a, feldker20a, weckesser21a, katz22a}.
Many ion-atom hybrid systems turned out to be chemically reactive which for the first time opened up possibilities to explore ion-neutral chemical processes in the mK to $\mu$K temperature regimes,
in particular charge exchange
\cite{schmid10a,hall11a,hall12a,doerfler19a,mills19a}.
Charge transfer can either occur radiatively, i.e., the collision system transits between different electronic states by the emission of a photon, or non-adiabatically around crossing points between two electronic states with a different asymptotic charge character \cite{tacconi11a, xing22a}. Both scenarios have previously been observed and characterized in different ion-atom hybrid systems, with non-adiabatic charge exchange usually being considerably more efficient than its radiative counterpart \cite{hall13a, xing22a}. Both non-reactive and reactive collisional processes in hybrid traps have often been successfully analyzed in the framework of Langevin or modified Langevin models only taking into account the long-range interactions between the ion and the neutral \cite{tomza19a, hall12a, hall13a,ratschbacher12a,meir16a}. 

A recent development is the introduction of molecular ions into hybrid experiments.
The exquisite degree of control over temperatures and quantum states of the trapped particles achievable in hybrid traps provide an ideal environment for studying subtle details of molecular phenomena and interactions \cite{hall12a,rellergert13a, puri17a,doerfler19a,hirzler21a}.
The increased complexity of molecular systems inevitably entails richer dynamics compared to atoms.
Previous studies on cold interactions of diatomic ions such as N$_2^+$ and O$_2^+$ with alkali atoms \cite{hall12a,doerfler19a} revealed an intricate interplay between long and short-range interactions governing the dynamics and kinetics of charge exchange in different electronic states of the collisions system.
In some states, charge transfer rates were found to be compatible with the Langevin model which only considers ion-induced dipole long-range interactions.
In other states, higher-order electrostatic interactions and, in particular, the specific dynamics around state crossings had to be taken into account in order to rationalize the charge transfer dynamics. 

In the present study, we introduced polyatomic molecular ions into an ion-atom hybrid experiment to explore the ramifications of molecular complexity in cold charge-transfer dynamics with the example of the prototypical triatomic species \ce{N2H+}, a stable linear cation well known from
astronomical observations
\cite{thaddeus_n2h+confirmation_1975},
laboratory experiments
\cite{saykally_n2h+MW_1976,gudeman_n2h+velocitymodulated_1983,Foster1984,sears_n2h+_1985}
and
theoretical works
\cite{Forsen1970,Brites2009,huang_comparison_2010}.
We studied the charge transfer of N$_2$H$^+$ with Rb atoms in their $(5s)~^2S_{1/2}$ ground and $(5p)~^2P_{3/2}$ excited electronic states at collision energies in the mK regime.
In both states, we found charge-transfer rates below the Langevin limit.
The experimental results were interpreted with the help of multireference \emph{ab-initio} calculations of the lowest potential energy surfaces (PESs) of the \ce{[N2H-Rb]+} collision system.
The calculations reveal that the charge transfer dynamics uncovered here is governed by distinct short-range molecular effects which cannot be captured by simple Langevin-type models and highlight the relevance of stereodynamics in cold collisions of this polyatomic system.

%%%%%%%%%%%%%%%%%%%%%%%%%%%%%%%%%%%%%%%%%%%%%%%%%%%%
\emph{Experiments.}
The experimental setup has been described in detail previous publications \cite{eberle16a,doerfler19a}. Briefly, it consisted of an ion-neutral hybrid trap comprising a linear-quadrupole radiofrequency (RF) ion trap embedded in a magneto-optical trap (MOT) for the cooling and trapping of Rb atoms. The ion trap was operated at a frequency of
$\Omega_\text{RF}=2\pi\times 3.25$~MHz with an amplitude of $V_\text{RF} = 400$~V and featured 12 individual electrode segments for the application of static and RF voltages. The experimental sequence started with the loading of Ca$^+$ ions produced from an effusive beam of Ca atoms by non-resonant photoionization. The Ca$^+$ ions formed Coulomb crystals \cite{willitsch12a} in the RF trap upon Doppler laser cooling on the $(4s)~^2S_{1/2}\leftrightarrow (4p)~^2P_{1/2} \leftrightarrow (3d)~^2D_{3/2}$ optical cycling transitions using two laser beams at 397 and 866~nm. N$_2^+$ molecular ions were produced by [2+1]-photon resonance enhanced photoionization from neutral nitrogen leaked into the vacuum chamber at a pressure of $\approx 1\times 10^{-8}$~mbar using a focused ns laser beam at 202~nm and sympathetically cooled into the center of the Ca$^+$ crystal \cite{hall12a, doerfler19a}. After the ionisation, the N$_2$ background pressure was maintained at that level for 30 seconds to equilibrate the rotational levels population by collisions and quench vibrational level populations to the ground state \cite{tong12a}. As demonstrated in Ref. \cite{kilaj18a}, the sympathetically cooled N$_2^+$ ions were converted into N$_2$H$^+$ by leaking hydrogen gas into the trap chamber for 30~s at a pressure of $\approx 1\times 10^{-8}$~mbar. 

\begin{figure}[!h]
\includegraphics[width=\columnwidth]{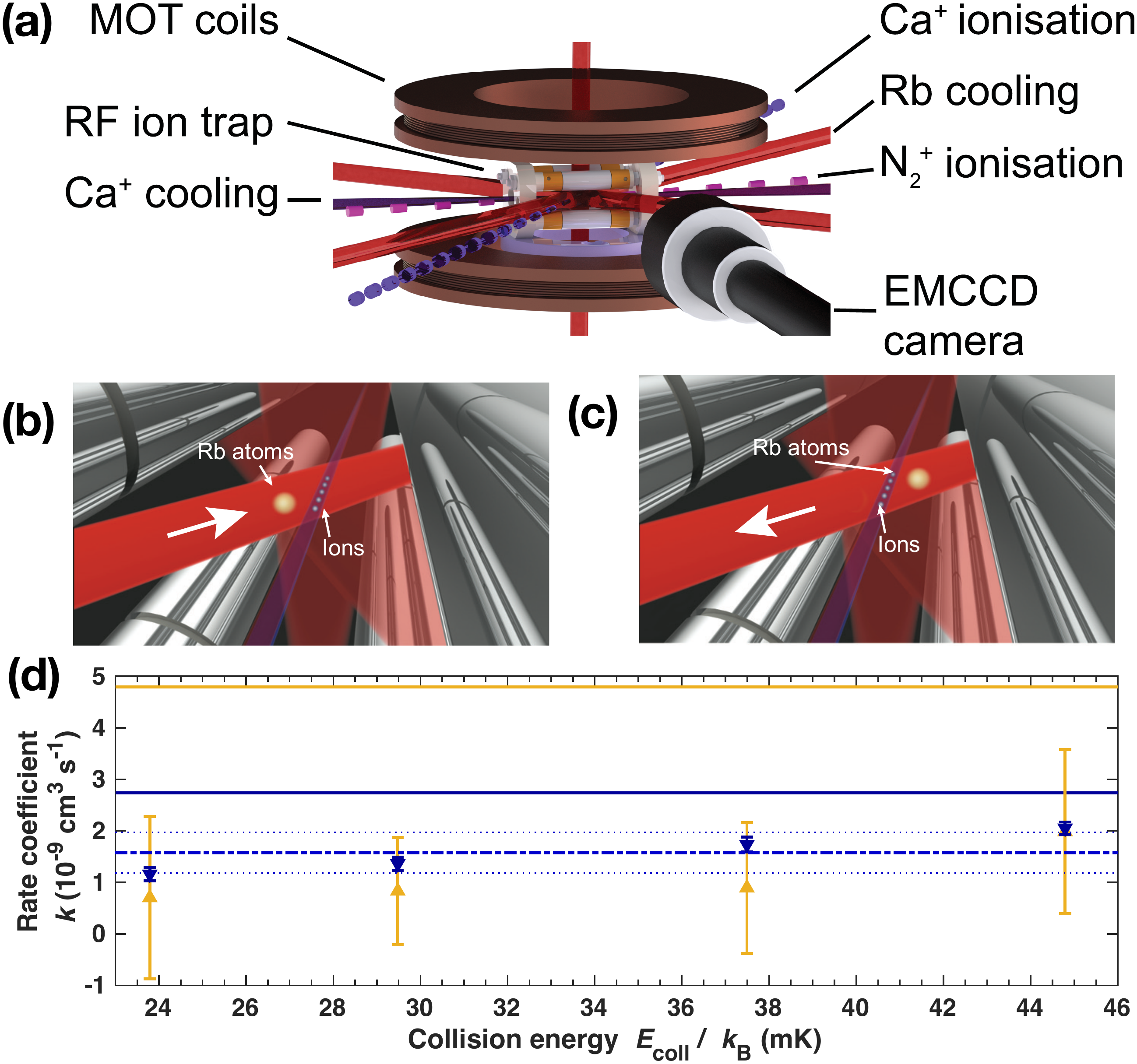}
\caption{\label{fig:expt}
(a) Schematic of the experiment consisting of a magneto-optical trap (MOT) for Rb atoms superimposed onto a radiofrequency (RF) linear-quadrupole ion trap. Laser beams are labeled by their functions. %Fluorescence images of atomic ions and atoms are recorded using an electron-multiplying charge-coupled device (EMCCD) camera.
(b) and (c) schematic for shuttling the cloud of ultracold Rb atoms in between two extremal positions through a Coulomb crystal of ions. (d) Charge-transfer rate coefficients as a function of the N$_2$H$^+$-Rb collision energy for the $(5s)~^2S_{1/2}$ ground state (blue \ding{116} symbols) and $(5p)~^2P_{3/2}$ excited state (orange \ding{115} symbols) of Rb. Blue (orange) solid line: Langevin rate constant in the Rb ground (excited) state. Dash-dotted and dashed lines: average experimental rate constant and its uncertainty limits. Error bars indicate the standard deviation of 5 measurements. }
\end{figure}

The MOT was continuously loaded from background Rb vapour replenished by an alkali-metal dispenser. Clouds of ultracold Rb atoms were generated in a position off-center from the ion trap (Fig. \ref{fig:expt} (b)). The Rb atoms were then accelerated using radiation-pressure forces to pass through the Coulomb crystal at well-defined velocities, recaptured after transit and sent back by applying radiation pressure in the opposite direction (Fig. \ref{fig:expt} (c)). Repetition of the sequence for varying magnitudes of radiation pressure enabled the study of ion-atom collisions at variable, well-defined collision velocities \cite{eberle16a}. Depending on whether the Rb cooling laser beams were switched off or on during transit, a part of the charge-transfer collisions between Rb and N$_2$H$^+$ occurred in the $(5p)~^2P_{3/2}$ excited state of Rb in addition the to the $(5s)~^2S_{1/2}$ ground state. The excitation fraction was determined from comparisons of the Rb fluorescence yields with an 8-level optical Block equation modelling as in Ref. \cite{hall13a}. Following Ref. \cite{doerfler19a}, rate constants for charge-exchange reactions between N$_2$H$^+$ and Rb were determined under pseudo-first-order-kinetics conditions for Rb by monitoring the depletion of the reactant ions from the crystal as a function of the time of exposure to the shuttling Rb atom cloud. The Rb density was extracted from a reference measurement of the O$_2^+$ + Rb reaction for which the rate coefficient conforms with the Langevin limit \cite{doerfler19a}.

Fig. \ref{fig:expt} (d) shows the charge-exchange rate coefficients, with Rb in either the $(5s)~^2S_{1/2}$ ground state or the $(5p)~^2P_{3/2}$ excited state, as functions of the collision energy in the interval $E_\text{coll}/k_\text{B}\approx 23-45$~mK. Within the uncertainty limits, the rate coefficients are identical for both states across the studied energy interval, in contrast to N$_2^+$ + Rb, in which a marked difference between the two channels was found \cite{doerfler19a}. The larger uncertainty of the data for the excited state is due to the error in the determination of the excited-state population \cite{doerfler19a}. The rate constant for reactions with Rb in the ground state seems to exhibit a slight positive energy dependence across the energy range studied. Neglecting this effect,
the average value is about a factor of 2 smaller than the corresponding Langevin rate coefficient, implying that the reaction rate is close to, but not equal to, the capture limit.

\emph{Theory.}
The reactants kinetic energy being nearly zero at temperatures of the order of few mK, endoenergetic reactions are impossible.
It is shown in \cref{fig:asymptotes} of the Supplemental Material (SM,
\footnote{%
\label{smfoot}%
See Supplemental Material for
the computation of asymptotic energies with methods described in
\cite{raghavachari_fifth-order_1989, hampel_comparison_1992,dunning1989,kendall1992,Hill2017} using MOLPRO \cite{werner_molpro_version,werner_molpro_2012,werner_molpro_2020};
excited state energies at the asymptotes estimated using data from \cite{johansson1961,sansonetti2006,miller_n2_1966,trickl_N2ion_1989,Yan2012a,Aymar2009};
computation of the \ce{N2H+} and \ce{N2H} potential energy curves with methods described in
\cite{werner1988,knowles1988,knowles1992,deskevich2004,werner1985,knowles1985,kreplin2019};
diabatic transition probabilities calculated using data from 
\cite{Foster1984};
relevant timescales estimated using data from
\cite{Foster1984,Amano2005}.
})
that the only energetically accessible product channels are -- aside from elastic or inelastic scattering -- (a) \ce{N2 + HRb+}, (b) \ce{N2Rb+ + H} and (c) \ce{N2 + H + Rb+}.
Any of those channels can be reached in their electronic ground states only.
Note that all those products involve an electron transfer from the rubidium atom to the hydrogen atom (even in the case of \ce{HRb+} the electron is essentially carried by \ce{H}, as it is more electronegative than \ce{Rb}).
The neutral radical \ce{N2H} is unstable and readily dissociates to give \ce{N2 + H},
as already shown in previous works \cite{Selgren1989,Mota2008,Talbi2007,Kashinski2012,Kashinski2015}
(the metastable bent geometry \cite{Bozkaya2010} is neglected here).
The \ce{N2Rb} radical is also unstable and can anyway not be observed as the channel \ce{N2 + Rb + H+} is energetically inaccessible.

\begin{figure}
	\includegraphics[width=3.375in]{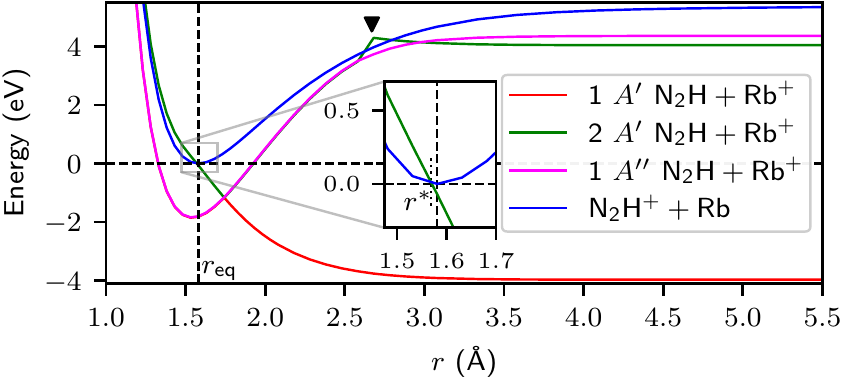}
	\caption{\label{fig:n2h-n2h+}
	Potential energy curves of linear \ce{N2H} and \ce{N2H+} at infinite separation from \ce{Rb+} and \ce{Rb}, respectively, as functions of the distance $r$ between the center-of-mass of \ce{N2} and \ce{H}.
	The \ce{N#N} bond distance is fixed at 1.0945~\AA{} and the dashed vertical line denotes the equilibrium value of $r$ at $r_{\text{eq}} = 1.5814$~\AA{} in \ce{N2H+} \cite{Brites2009}.
    %The solid blue curve is drawn by shifting the dashed blue one, obtained from the calculations, by the ionization energy of \ce{Rb}.
 	Energies are relative to the threshold which corresponds to the minimum of the \ce{N2H+ + Rb} potential (solid blue curve).
	\ding{116}: see computational methods in the SM [34].
    }
\end{figure}

\cref{fig:n2h-n2h+} shows the computed potential energy curves (PECs) of the linear \ce{N2H} and \ce{N2H+} as functions of the distance $r$ between H and the center of mass of \ce{N2} at fixed \ce{N#N} distance (methods are discussed in the SM [34]).
These are corrected for the ionization energy of \ce{Rb} so that they mimic the curves of \ce{N2H + Rb+} and \ce{N2H+ + Rb}, respectively, at infinite atom-cation separation.
The \ce{N2H+ ({+} Rb)} potential is that of a bound species.
The potential energy landscape of \ce{N2H ({+} Rb+)} is more complex,
featuring one steep dissociative state which crosses a pair of degenerate states,
as already shown in Ref.~\cite{Mota2008}.
The \ce{N2H ({+} Rb+)} dissociative state plays the major role in the electron transfer process, as its PEC intersects that of \ce{N2H+ ({+} Rb)}.
This intersection happens to be located at $r^\ast$
slightly shorter than $r_{\text{eq}}=1.5814$~\AA{}, the \ce{N2-H} distance at the equilibrium geometry of \ce{N2H+},
and drives the charge transfer \ce{N2H+ + Rb -> N2H + Rb+} as the \ce{Rb} atom approaches \ce{N2H+}.

\begin{figure}
    \centering
    \includegraphics[width=\columnwidth]{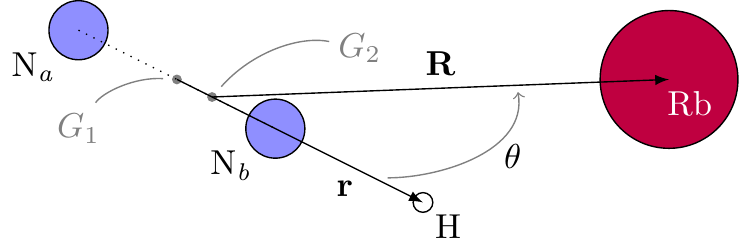}
    \caption{Definition of the coordinates describing the arrangement of the nuclei in the \ce{[N2H-Rb]+} system.
    The Jacobi vectors $\textbf{r}$ and $\textbf{R}$ are such that $\textbf{r}$ points from the center-of-mass of \ce{N2} ($G_1$) to \ce{H} and $\textbf{R}$ points from the center-of-mass of \ce{N2H} ($G_2$) to \ce{Rb}.
    The \ce{N#N} bond length is kept fixed and \ce{H} stays along the \ce{N-N} axis, as in \cref{fig:n2h-n2h+}.
    Within these constraints, the geometry of the system is fully determined by the distances $r = |\textbf{r}|$, $R = |\textbf{R}|$, and the angle $\theta$ between the two vectors.}
    \label{fig:geometry}
\end{figure}

\begin{figure*}
    \includegraphics{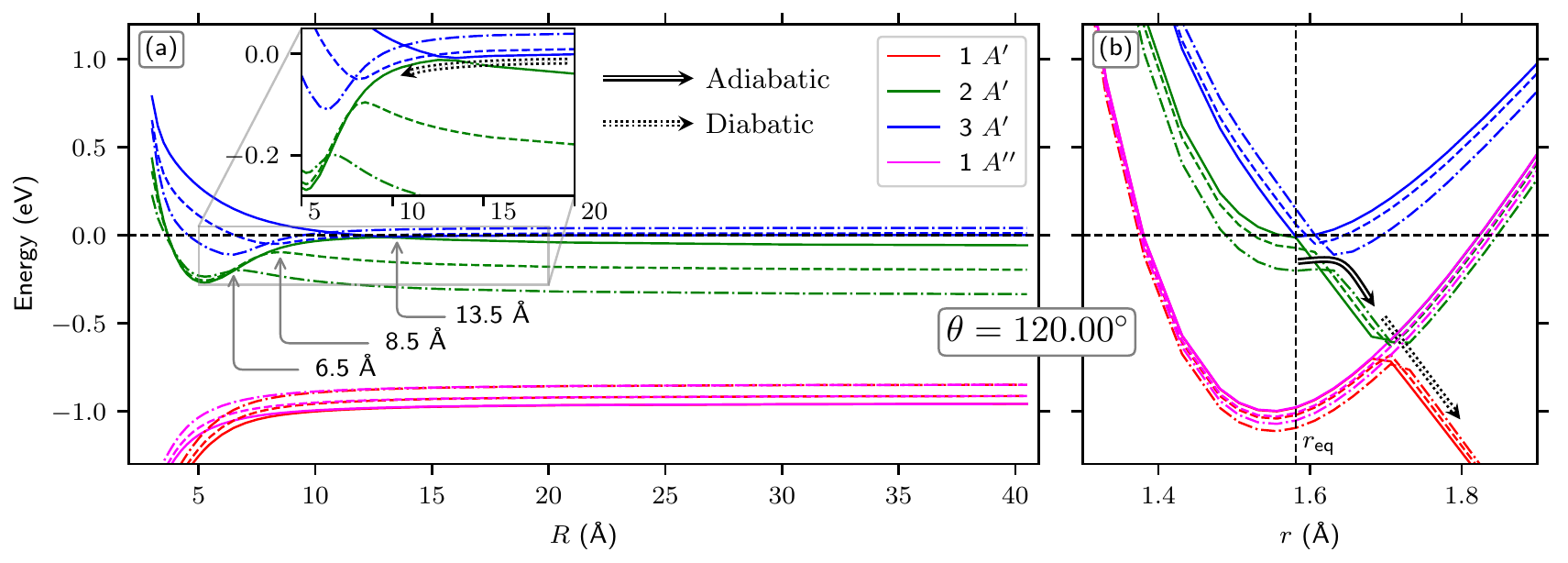}%
    \caption{
    \label{fig:cuts-1d}
    Adiabatic potential energy surfaces cuts of the \ce{N2H+ + Rb} system as functions of (a)
    the atom-cation distance $R$ and (b) the \ce{N2-H} distance $r$, for $\theta=120^{\circ}$.
    Note that, in panel~(b), the state labeling/coloring is reversed compared to \cref{fig:n2h-n2h+} for $r$ shorter than $r^\ast$ since here the electronic eigenstates are ordered by increasing energy (within a given irreducible representation).
    In (a)
    solid lines
    (\protect\tikz[baseline=-0.5ex]{ \protect\draw (0,0) -- (0.5,0); })
    correspond to cuts at $r=r_{\text{eq}}=1.5814$~\AA{},
    dashed lines
    (\protect\tikz[baseline=-0.5ex]{ \protect\draw[densely dashed] (0,0) -- (0.45,0); })
    at $r=r_{\text{eq}}+0.025$~\AA{}
    and dash-dotted lines
    (\protect\tikz[baseline=-0.5ex]{ \protect\draw[densely dashdotted] (0,0) -- (0.5,0); })
    at $r=r_{\text{eq}}+0.050$~\AA{}.
    The indicated values of $R$ locate the positions of the avoided crossings along those cuts.
    In (b),
    solid lines
    (\protect\tikz[baseline=-0.5ex]{ \protect\draw (0,0) -- (0.5,0); }) correspond to cuts at $R=13.5$~\AA{},
    dashed lines
    (\protect\tikz[baseline=-0.5ex]{ \protect\draw[densely dashed] (0,0) -- (0.45,0); })
    at $R=8.5$~\AA{}
    and dash-dotted lines
    (\protect\tikz[baseline=-0.5ex]{ \protect\draw[densely dashdotted] (0,0) -- (0.5,0); })
    at $R=6.5$~\AA{}.
    The dashed horizontal black lines denote the entrance channel energy.
    Double arrows indicate transitions described in the text.
    }
\end{figure*}

Thus, we computed the four lowest electronic eigenstates of the full tetra-atomic \ce{[N2H-Rb]+} system identified in \cref{fig:n2h-n2h+} as functions of the three coordinates $r$, $R$ and $\theta$ defined in \cref{fig:geometry}.
\cref{fig:cuts-1d} displays cuts of the corresponding PESs for $\theta=120^{\circ}$ (for other angles, see SM [34]).
The system is initially in the diabatic \ce{N2H+ + Rb} state at ($r=r_\text{eq},R\to\infty$).
This corresponds to the solid blue curve at large $R$ in \cref{fig:cuts-1d}~(a) and the bottom of the curve obtained by joining those of state $2A'$ in green for $r < r^\ast$ and $3A'$ in blue for $r > r^\ast$ in \cref{fig:cuts-1d}~(b).
The intersection point at $r^\ast$ shown in \cref{fig:n2h-n2h+} is in fact an avoided crossing between the PECs in \cref{fig:cuts-1d}~(b) whose energy gap is exactly zero at $R\to\infty$.

For \ce{N2H+} in its equilibrium geometry (solid curves in \cref{fig:cuts-1d} (a)), the adiabatic PECs of states $3A'$ and $2A'$ feature an avoided crossing around $R=13.5$~\AA{} (see inset).
There, the energy spacing between these states is still nearly zero.
On the other hand, the attractive interaction between \ce{N2H+} and \ce{Rb} is such that the full system acquires enough kinetic energy to undergo a diabatic transition $3A' \to 2A'$ with almost unit probability (dotted arrow in the inset of \cref{fig:cuts-1d}(a), see also Landau-Zener transition probabilities estimates in \cref{tab:lz} in the SM [34]).
However, for shorter distances $R$,
the position of the avoided crossing varies in the vicinity of $r_\text{eq}$ (see \cref{fig:cuts-1d} (b))
and remains at energies below the entrance channel threshold.
There, the energy gap between the two adiabatic states widens.
Thus, for shorter $R$, diabatic transitions between the states $3A'$ and $2A'$ are less effective, giving chances to the system to stay adiabatically in the $2A'$ state along the $r$ coordinate (see \cref{tab:lz} in the SM [34]).
In other words, the system can leak from the parabolic part of the green PEC around $r_\text{eq}$, with \ce{N2H+ + Rb} character, to the steep portion at $r>r^\ast$ where the system gains \ce{N2H + Rb+} character (solid arrow in \cref{fig:cuts-1d} (b)).
\emph{This adiabatic passage is the manifestation of the electron transfer between \ce{Rb} and \ce{N2H+} which is promoted by a deformation of the molecule along the $r$ coordinate.}
The Landau-Zener model roughly predicts a 5\% chance per pass to remain adiabatically in state $2A'$.
However, the typical time spent by \ce{Rb} in the vicinity of the cation ($\approx 2.3$~ps) is two orders of magnitude longer than the \ce{N2-H+} stretching vibrational period ($\approx 15$~fs).
Thus, the system has enough time to leak through the avoided crossing via this vibration (see SM [34] for timescales estimates).
Once on the dissociative part of the $2A'$ state PEC (green curves, $r>r^\ast$), the system diabatically transits to its continuation on state $1A'$ (see dotted arrow in \cref{fig:cuts-1d} (b)),
i.e. \ce{H} readily dissociates from \ce{N2} as a result of the electron transfer from \ce{Rb} to \ce{N2H+}.
% In summary, the coupling of the collision coordinate $R$ to the N$_2$H-internal coordinate $r$ is essential for the charge transfer to occur.
\begin{figure}
    \includegraphics{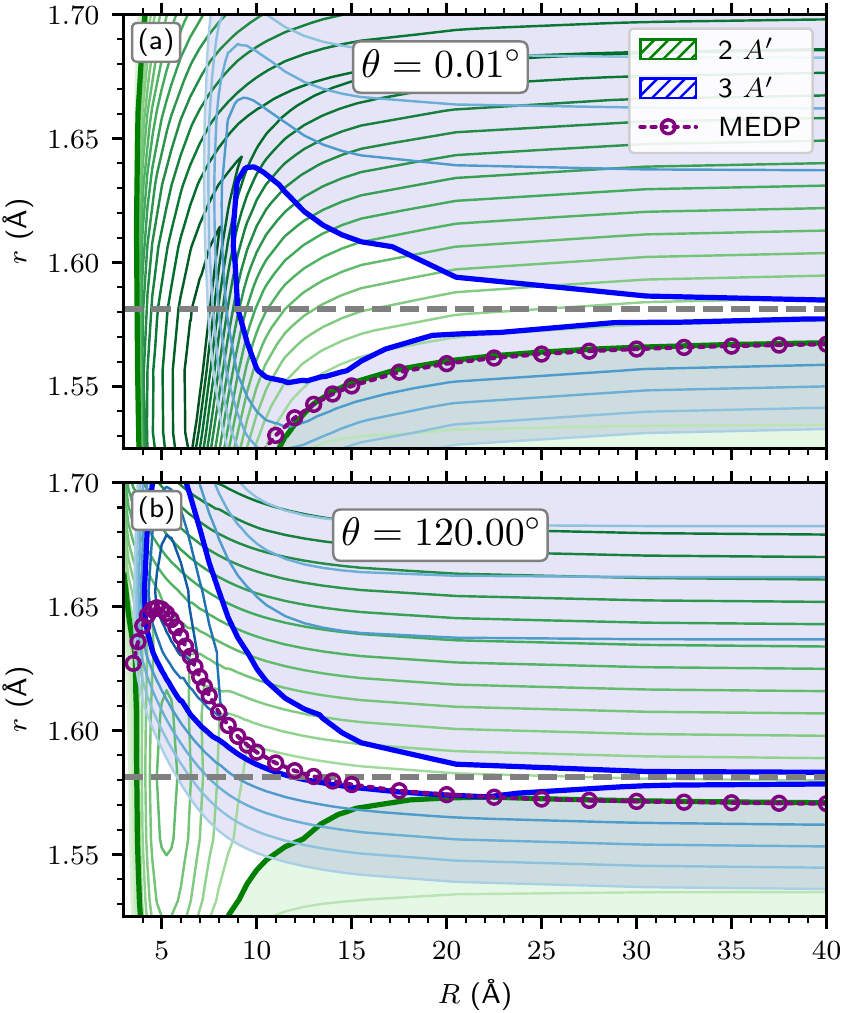}
    \caption{Contour plots of the PESs of states $2A'$ and $3A'$ as functions of $r$ and $R$ for
    (a) $\theta = 0.01^\circ$
    and
    (b) $\theta = 120^{\circ}$.
    The thick green and blue contour lines correspond to the threshold energy $E=0$ in those states, respectively.
    Other isoenergetic contours are traced every 0.05~eV.
    Shaded areas indicate energy-forbidden regions of configuration space on the corresponding PES.
    The dashed grey line indicates the entrance asymptotic channel $(r=r_\text{eq},R\to\infty)$.
    }
    \label{fig:path-contour}
\end{figure}

The discrepancy between the experimentally measured reaction rates and
those predicted by the Langevin capture model is attributed to
specific features of the PESs which are dependent on the angle of attack $\theta$.
\cref{fig:path-contour} shows two-dimensional contour plots in $r$ and $R$ of the PESs of states $2A'$ and $3A'$ cut at $\theta=0.01^\circ$
and $\theta=120^\circ$ (for other angles, see SM [34]).
For each angle, the seam of avoided crossings from asymptotic separation towards finite distance $R$ is traced by a minimum energy difference path (MEDP).
As \cref{fig:path-contour} shows, the PESs display a strong anisotropy at short range.
Starting from the asymptote $(r,R)\to(r_\text{eq},+\infty)$ on state $3A'$,
when the collision takes place on the \ce{H}--side of \ce{N2H+} ($\theta\approx 0^{\circ}$, \cref{fig:path-contour}~(a)), the region of coordinate space accessible by the system energy-wise is limited to distances $R$ no shorter than approximately 9~\AA{}.
In this region, the PES of state $3A'$ is very shallow (the minimum is about 0.034~eV below the threshold) and the MEDP never crosses its boundary.
The PES of state $2A'$ underneath steeply decreases in energy both with increasing $R$ and $r$, eventually becoming -- in a diabatic sense -- state $1A'$ (see \cref{fig:cuts-1d} (b)).
The limited extent of the energy-allowed region and the increase of the energy gap between the states $2A'$ and $3A'$ therein forbids any non-adiabatic transition, hence no reaction takes place.

Conversely, when the collision occurs at $\theta=120^\circ$, i.e. \ce{Rb} approaches \ce{N2H+} non-collinearly from the \ce{N2}--side (\cref{fig:path-contour} (b)), the energy-allowed region of the PES of state $3A'$ reaches points with $R$ as close as 4.2~\AA{}.
The lowest points in energy in this PES cut ($\approx 0.13$~eV below the threshold) are close to the MEDP which, in this case, clearly enters inside that region.
There, the energy gap between the PESs of states $3A'$ and $2A'$ is much smaller than in the $\theta=0^\circ$ case (see also \cref{fig:si:path120} in SM [34]).
Hence, the latter state is more likely reached by non-adiabatic transition and the reaction occurs as discussed earlier in connection with \cref{fig:cuts-1d}.
The case $\theta = 60^\circ$ (see \cref{fig:si:path60} in the SM [34]), where the MEDP is nearly tangent to the energy-allowed region boundary, is close to the limit between the two behaviors described above.

\emph{Conclusion.}
Unaware of the structure of the cation, the Langevin capture model considers only the isotropic long-range ion-atom interaction,
which is satisfactory when no other hindrance to the charge exchange arises at short range.
However, our inspection of the distinctive interaction between \ce{N2H+} and \ce{Rb} at short range,
necessarily anisotropic, reveals that an efficient charge transfer in the cold regime cannot occur from all angles of attack because of the above-mentioned obstacles encountered for $\theta<60^{\circ}$, i.e. for about one third of the configuration space spanned by $\theta$.
Although the orientation of the cation is not locked during the collision as \cref{fig:path-contour} may suggest (see SM [34])
these obstacles are nonetheless a limitation to a successful reaction and are qualitatively reflected in the experimental reaction rates which are roughly half of the Langevin prediction.

The anisotropy of the potentials also precludes the formation of the \ce{HRb+} product.
From a steric point of view, the proton capture by \ce{Rb} should be eased by an attack around $\theta=0^\circ$,
but no effective electronic transition allows the formation of the \ce{H\cdots Rb+} bond
leaving no other option than scattering of \ce{Rb} away from \ce{N2H+}.
Conversely, the attack of \ce{Rb} from the \ce{N2}--side does allow the electron transfer, leading to the formation of the ephemeral \ce{N2H} radical.
Due to the steep dissociative character of the PES of state $2A'$ (and its diabatic continuation in state $1A'$), the radical readily dissociates into \ce{N2 + H}.

\emph{Summary.} We have studied the charge exchange of a polyatomic ion with Rb in an ion-atom hybrid trap. At the present mK collision energies, the system was found to only sample parts of the available configuration space of the reaction so that charge transfer becomes highly geometry dependent, is promoted by a deformation of the molecule and thus becomes slower than the Langevin limit. The present study underlines the role of stereodynamics and polyatomic effects in cold collisions of molecular ion-neutral systems.

%TC:ignore

\begin{acknowledgments}
\emph{Acknowledgments.} AV thanks T.~V\' ery and R.~Lacroix for extensive support in the usage of the Jean-Zay CNRS-IDRIS supercomputer. AV thanks Amrendra Pandey (Laboratoire Aim\'e Cotton) for fruitful discussions at the early stage of the computational work and the \textsc{Molpro} development team for technical assistance. LW thanks J. Brand\~ao (U. Algarve, PT) for many early discussions.
This work is supported by Investissements d'Avenir, LabEx PALM (ANR-10-LABX-0039-PALM), Universit\'e Paris-Saclay, the Swiss National Science Foundation (grants nr. 200020\_175533 and TMAG-2\_209193) and the University of Basel.
Computation on the Jean-Zay CNRS-IDRIS supercomputer was made possible thanks to contract A0120810769.

AV and AD contributed equally to this work. % see: https://journals.aps.org/prl/authors/byline-addresses-footnotes-acknowledgments-statements-about-authors-h22
\end{acknowledgments}

\bibliography{biblio}

%TC:endignore
\end{document}